\documentclass[a4paper]{jpconf}

\usepackage{graphicx}

\newcommand{\p}{\prime}
\newcommand{\sumi}{\sum_{i=0,\,\mathrm{even}}^{n-1}}

\begin{document}
\title{Generalized Parton Distributions from Lattice QCD}
\author{Dru B.\ Renner}
\address{Department of Physics, University of Arizona, 1118 E 4th Street, Tucson, AZ 85721, USA}
\begin{abstract}
%
%
%
%
I review the LHPC Collaboration's lattice QCD calculations of the generalized parton distributions of the nucleon
and highlight those aspects of nucleon structure best illuminated by lattice QCD,
the nucleon's spin decomposition and transverse quark structure.
\end{abstract}
\section{Introduction}
Generalized parton distributions provide the means to calculate aspects of nucleon
structure not previously accessible to lattice QCD calculations.  Primary among
these are the nucleon spin decomposition~\cite{Ji:1996ek} and the transverse
quark structure~\cite{Burkardt:2000za}.  The generalized parton distributions
determine the fraction of nucleon spin carried by quark helicity, quark orbital
motion, and gluons.
Additionally, the generalized parton distributions
provide a three dimensional picture of a fast moving nucleon by extending 
the ordinary parton distributions into the transverse plane.
%
%

The Lattice Hadron Physics Collaboration is pursuing a program to calculate the
nucleon's generalized parton distributions.  Initial calculations have been 
completed for pion masses from $750~\mathrm{MeV}$ to $900~\mathrm{MeV}$~\cite{Hagler:2003jd,:2003is}.
Additionally current calculations are underway covering the region of pion masses from
$350~\mathrm{MeV}$ to $750~\mathrm{MeV}$~\cite{Renner:2004ck}.  In these
proceedings I review the results from our earlier calculations highlighting the 
potential impact of lattice QCD calculations on our understanding of quark orbital 
motion and transverse quark structure within the nucleon.
\section{Generalized Form Factors}
\label{gen}
Definitions of the generalized parton distributions can be found 
in~\cite{Diehl:2003ny}.  However, there is an equivalent and more convenient
language which is naturally suited to lattice calculations, generalized form factors.
The generalized form factors are defined in terms of matrix elements of the
twist two operators which can be calculated on the lattice.  For example consider
the tower of twist two operators 
$O_q^{\mu_1 \cdots \mu_n} = \overline{q} iD^{(\mu_1} \cdots iD^{\mu_{n-1}} \gamma^{\mu_n)} q$.
There is a corresponding tower of generalized form factors, $A^q_{ni}$, $B^q_{ni}$, and $C^q_{n}$, defined by 
\begin{equation}
\label{gffs}
\langle P^\p | O_q^{\mu_1 \ldots \mu_n} | P \rangle = 
\overline{U}(P^\p) [ \sumi \left( A^q_{ni}(t) K^A_{ni} + B^q_{ni}(t) K^B_{ni} \right)
                              + \delta^n_\mathrm{even} C^q_n(t)    K^C_{n}\,\, ] U(P)
\end{equation}
where the $K$ are kinematic functions of $P$ and $P^\p$ and
the spin labels have been suppressed.
\section{Lattice Calculation}
\label{latt}
The details of our lattice calculation are given in~\cite{Dolgov:2002zm} and \cite{Hagler:2003jd}.  In brief, we
calculate the matrix elements on the left hand side of equation~\ref{gffs} on the lattice.
We then match these results to the $\overline{MS}$ scheme at $\mu=2~\mathrm{GeV}$, and invert
equation~\ref{gffs} to determine the generalized form factors.  A key feature of our method
is that we calculate an exhaustive set of matrix elements and solve the resulting overdetermined
system of equations to extract the generalized form factors with as small an error as 
possible.  In the following I use only results from a single lattice calculation with 
$m_\pi=896(6)~\mathrm{MeV}$.  Further details can be found in the references above.  Additionally,
there are other similar calculations in the literature~\cite{Gockeler:2003jfa,Gockeler:2003jfb,Gockeler:2004vx,Gockeler:2004mn,Mathur:1999uf}.
\section{Quark Orbital Motion}
\label{spin}
The nucleon spin can be written as the sum of three separately gauge invariant observables,
\begin{equation}
\label{decomp}
1/2=1/2\,\Delta\Sigma^{u+d} +L^{u+d} +J^g,
\end{equation}
where the terms from left to right are the contributions from the quark helicity, quark 
orbital angular momentum, and total gluon angular momentum.  The observables in equation~\ref{decomp}
are each given by singlet matrix elements which pose a significant challenge to lattice
calculations.  Singlet matrix elements require the computation of disconnected quark 
diagrams which are computationally quite demanding.
%

\begin{figure}[t]
\begin{minipage}{18pc}
\includegraphics[width=18pc]{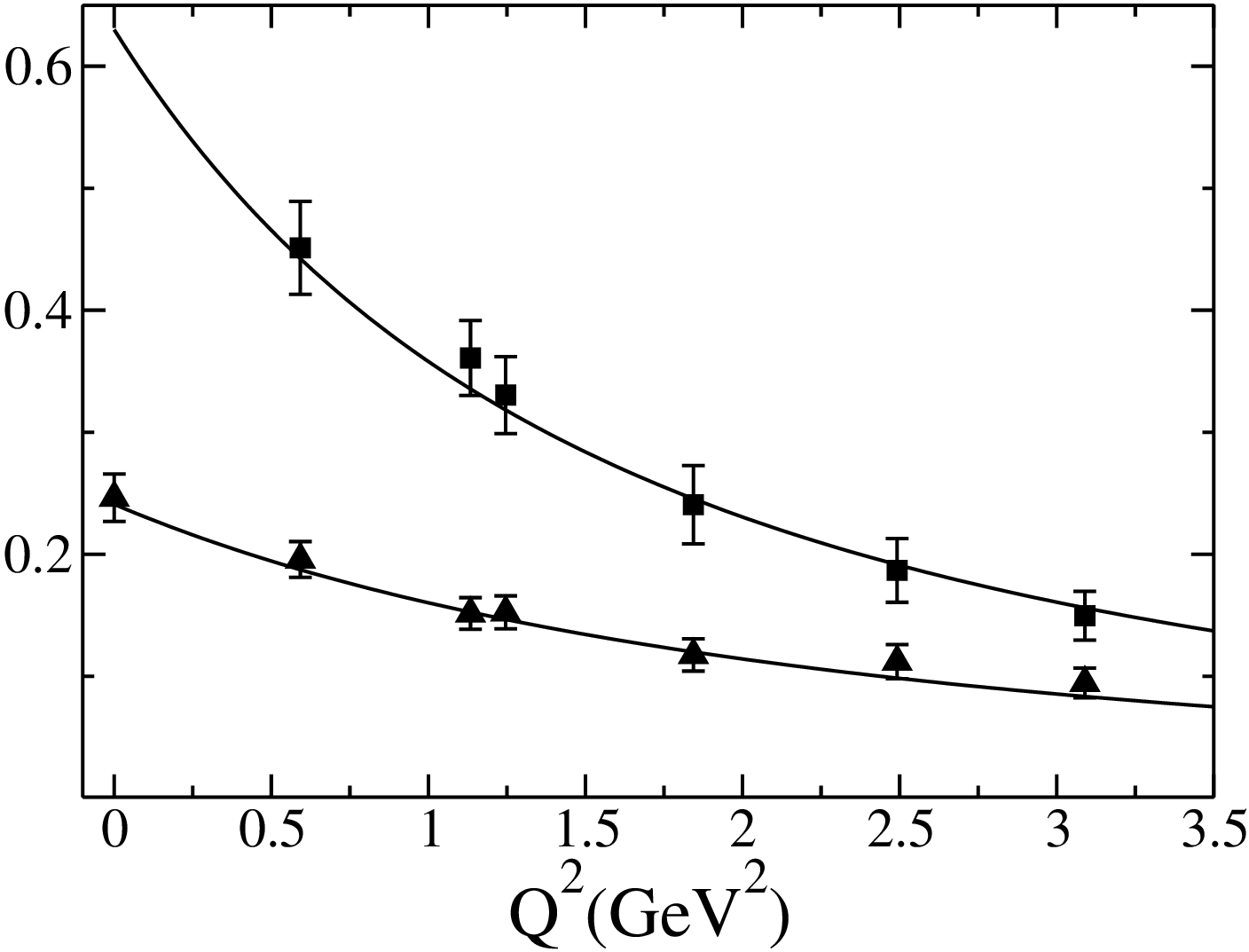}
\caption{\label{ABC}squares are $B^{u-d}_{20}$, triangles are $A^{u-d}_{20}$}
\end{minipage}\hspace{2pc}
\begin{minipage}{18pc}
\includegraphics[width=18pc]{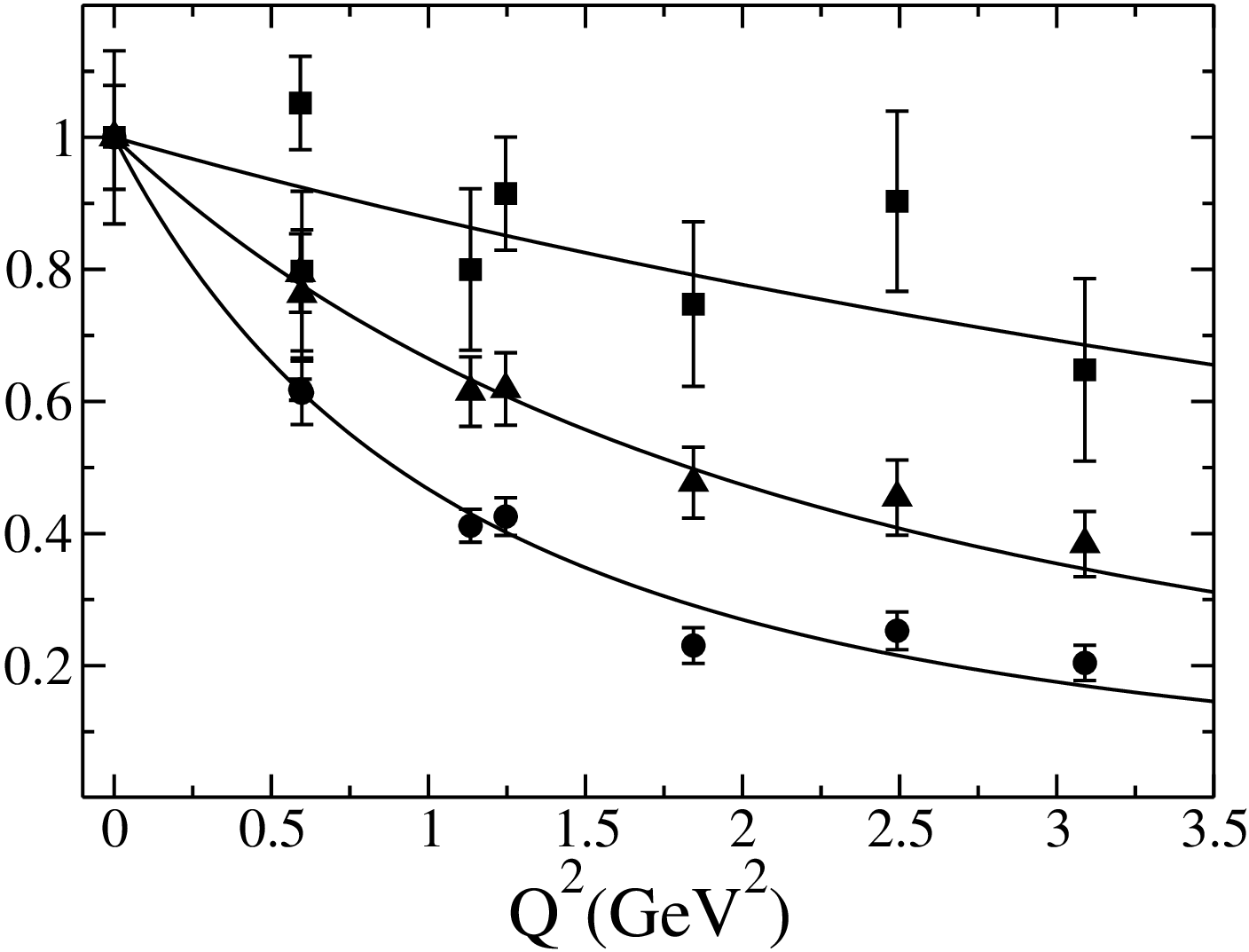}
\caption{\label{AAA}squares are $A^{u-d}_{30}$, triangles are $A^{u-d}_{20}$, circles are $A^{u-d}_{10}$}
\end{minipage} 
\end{figure}

However non-singlet matrix elements receive no contributions from disconnected
diagrams and are readily calculated on the lattice.
In our calculation~\cite{Negele:2003ma}, shown in figure~\ref{ABC}, we found that $L^{u-d}=-0.193(32)$
indicating that at least one of the quark flavors is undergoing significant
orbital motion.  This is quite interesting given that the connected 
contribution to $L^{u+d}$ is $0.002(3)$ which, ignoring disconnected diagrams,
suggests a strong cancellation among the quark flavors.
Due to the missing contributions to $L^{u+d}$ we are incapable of determining 
$L^u$ and $L^d$ separately, however, we can set a lower bound on the typical 
contribution from the quark orbital motion: the average magnitude of the
quark orbital angular momentum, $L_{\mathrm{ave}} = 1/2(|L^u|+|L^d|)$,
is bounded from below, $L_{\mathrm{ave}} \ge |L^{u-d}|$, indicating that on average
the quarks are contributing, by way of their orbital motion, on 
the order of $20\%$ of the nucleon's total spin.
%
%
\section{Transverse Quark Structure}
\label{tran}
The transverse quark distribution, $q(x,\vec{b}_\perp)$, gives 
the probability to find a quark $q$ with a momentum fraction $x$ at a distance 
$\vec{b}_\perp$ from the transverse center of the nucleon.  The $x$ moments 
of the transverse distribution are given by
\begin{equation}
\label{mom}
q_n(\vec{b}_\perp) = \int_{-1}^{1}\!dx\,\, x^{n-1}\, q(x,\vec{b}_\perp) = \int \frac{d^2\Delta_\perp}{(2\pi)^2} e^{-i\vec{b}_\perp\cdot\vec{\Delta}_\perp} A^q_{n0}(-\vec{\Delta}_\perp^2).
\end{equation}
The $Q^2$ dependence of $A^q_{n0}$ is calculated directly on the lattice and, as explained shortly,
for large $n$ probes the transverse distribution of quarks with $x$ near $1$.  
Specifically we calculate the transverse rms radius and average $x$ of the distribution $\int\!dx\,x^{n-1} q(x,\vec{b}_\perp)$ allowing
us to indirectly examine the $x$ dependence of the transverse size of the nucleon.  Additionally, upon 
entertaining a few assumptions, we can construct the $\vec{b}_\perp$ dependence of $q(x,\vec{b}_\perp)$
as well.
The details of the following calculations are given in~\cite{Schroers:2003mf} and \cite{:2003is}.
\subsection{$Q^2$ Dependence}
\label{qdep}
The $Q^2$ dependence of, and in particular the slope of, $A^q_{n0}$ determines the rms radius of the $n^\mathrm{th}$ moment of $q(x,\vec{b}_\perp)$,
\begin{equation}
\label{rmseq}
\left< b_{\perp}^2 \right>_{n} =
\frac{ \int\!d^2b_{\perp}\, b_{\perp}^2\, \int_{-1}^{1}\! dx\, x^{n-1}q(x,\vec{b}_{\perp}) }
{ \int\!d^2b_{\perp}\, \int_{-1}^{1}\! dx\, x^{n-1}q(x,\vec{b}_{\perp}) }
=\frac{-4}{A^q_{n0}(0)}\frac{dA^q_{n0}(0)}{dQ^2}.
\end{equation}
Notice that for large $n$ the integrals in equation~\ref{rmseq} are dominated by the 
limit $x\rightarrow 1$.  Furthermore, as $x\rightarrow 1$ a single quark carries all
the longitudinal momentum of the nucleon.  As discussed in~\cite{:2003is},
a quark in this limit is kinematically constrained to have $\vec{b}_\perp\rightarrow 0$.
Additionally, explicit light cone wave functions show $q(x,\vec{b}_\perp)$ behaves as $\delta^2(\vec{b}_\perp)$
for $x\rightarrow 1$.  Thus equation~\ref{rmseq} demonstrates that the slope, and corresponding 
rms radius, of $A^q_{n0}$ vanishes as $n\rightarrow\infty$.  It is quite reassuring that 
we find, as shown in figure~\ref{AAA}, that even for $n=1$, $2$, and $3$ we see the slopes 
of $A^q_{n0}$ decreasing with increasing $n$ indicating that the transverse distribution of
quarks within the nucleon is indeed becoming more narrow as $x$ approaches $1$. 
\subsection{$x$ Dependence}
\label{xdep}
\begin{figure}[t]
\begin{minipage}{18pc}
\includegraphics[width=18pc]{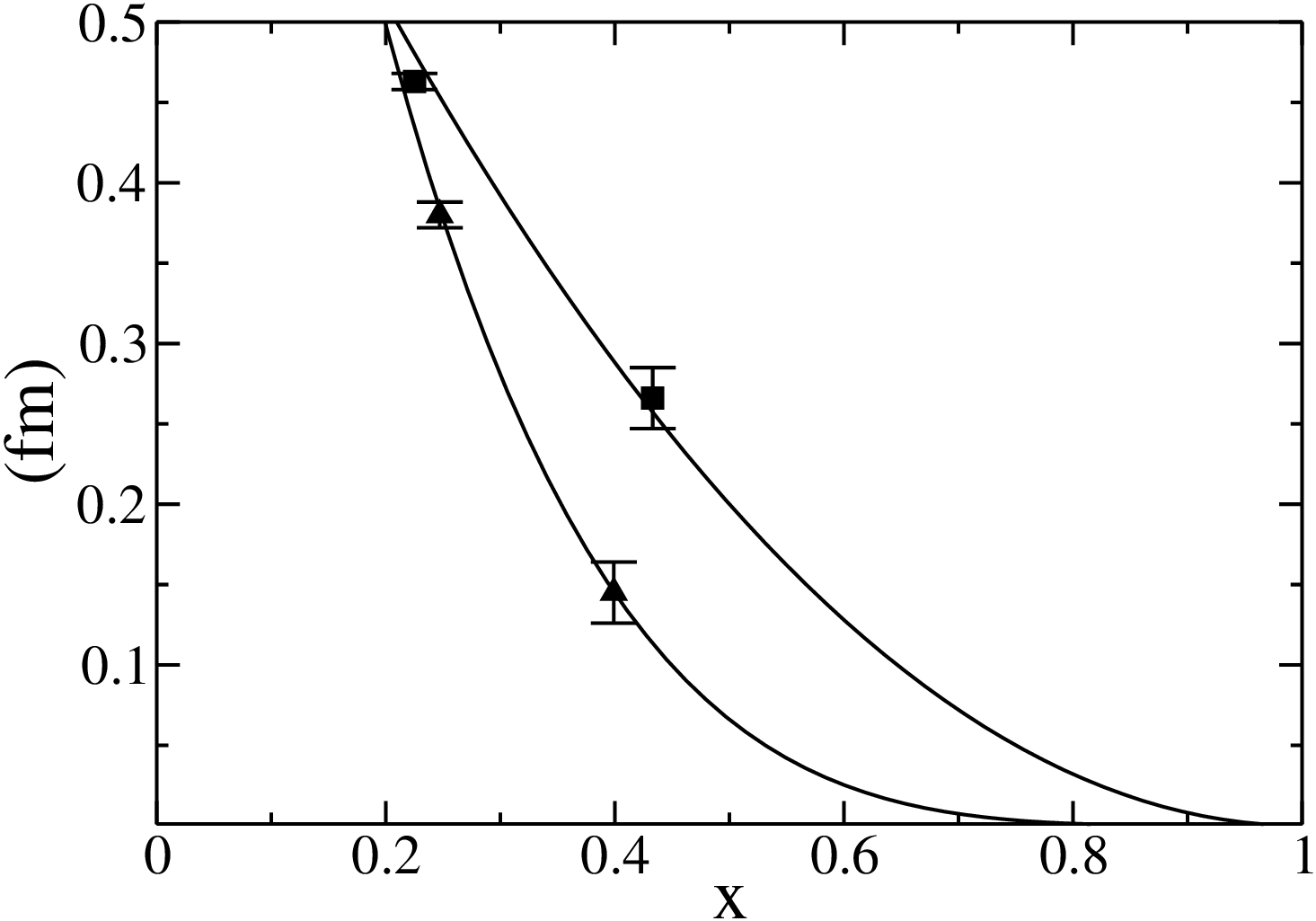}
\caption{\label{rms}squares are $\langle b^2_\perp \rangle^{u+d}_{(n)}$, triangles are $\langle b^2_\perp \rangle^{u-d}_{(n)}$ for $n=1,3$}
\end{minipage}\hspace{2pc}
\begin{minipage}{18pc}
\includegraphics[width=18pc]{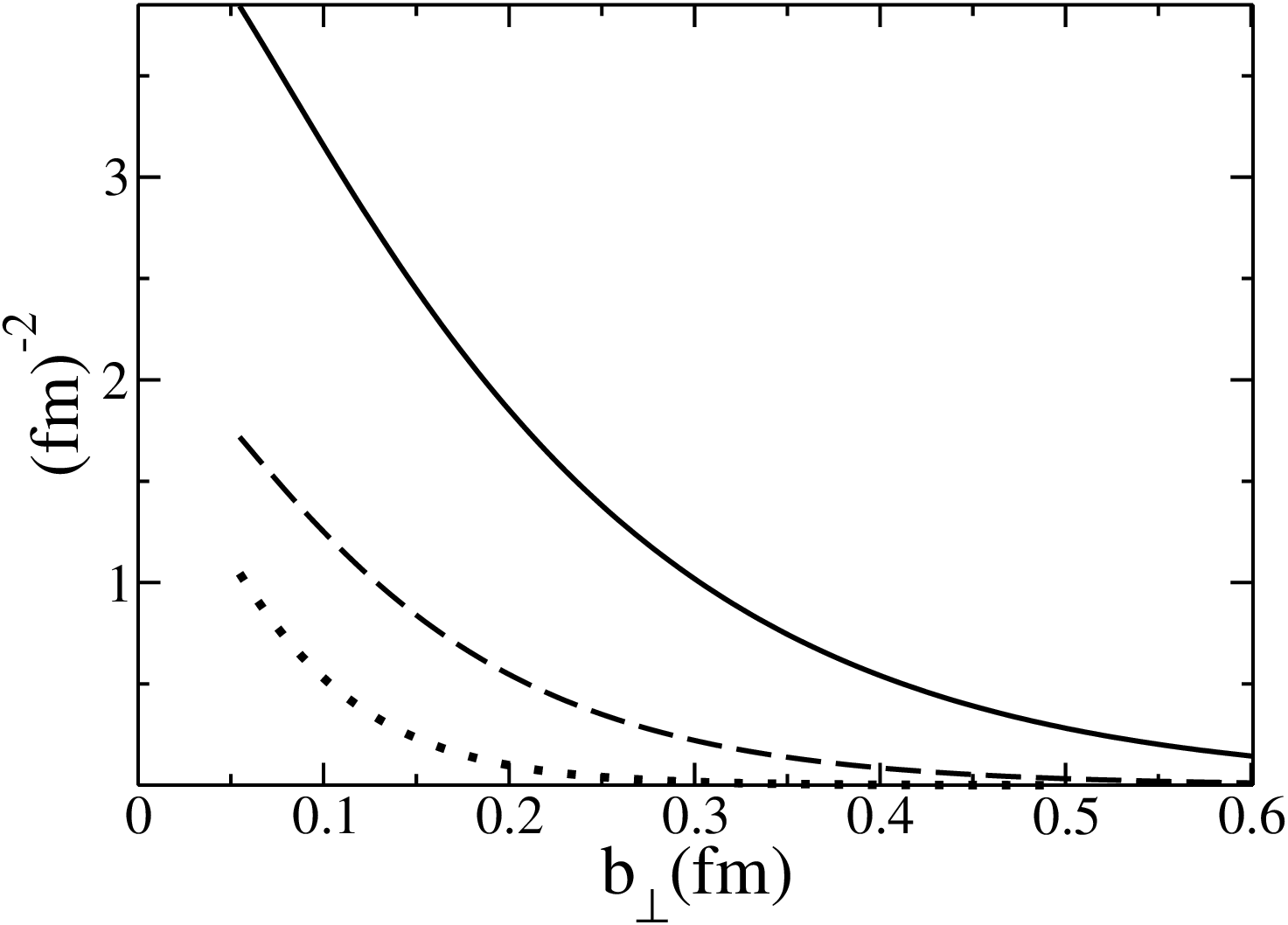}
\caption{\label{qqq}solid, dashed, dotted are $q_n(\vec{b}_\perp)$ for $n=1,2,3$, $q=u-d$}
\end{minipage} 
\end{figure}

The rms radii $\left< b_{\perp}^2 \right>_n$ in equation~\ref{rmseq} give the transverse size at a fixed 
moment $n$ rather than at a fixed momentum fraction $x$,
\begin{displaymath}
\left< b_{\perp}^2 \right>_x =
\frac{ \int\!d^2b_{\perp}\, b_{\perp}^2\, q(x,\vec{b}_{\perp}) }
{ \int\!d^2b_{\perp}\, q(x,\vec{b}_{\perp}) }.
\end{displaymath}
However for a fixed $n$ there is a region of $x$ which dominates $\left< b_{\perp}^2 \right>_n$,
thus we can think of $\left< b_{\perp}^2 \right>_n$ as an average of $\left< b_{\perp}^2 \right>_x$
over a region centered on the average $x$ of the $n^\mathrm{th}$ moment,
\begin{equation}
\label{xave}
\left<x\right>_n =
\frac{ \int\!d^2b_{\perp}\, \int_{-1}^{1}\! dx\, \left|x\right| x^{n-1}q(x,\vec{b}_{\perp}) }
{ \int\!d^2b_{\perp}\, \int_{-1}^{1}\! dx\, x^{n-1}q(x,\vec{b}_{\perp}) }
= \frac{<x^n>+2(-1)^n\int\!d^2b_{\perp}\int_{0}^{1}\!dx\,x^n\overline{q}(x)}{<x^{n-1}>} \approx \frac{<x^n>}{<x^{n-1}>}.
\end{equation}
The anti-quark 
contribution in equation~\ref{xave} is not accessible to current lattice calculations, however we 
expect it to be small and simple guesses based on the phenomenologically determined distributions 
indicated the following qualitative conclusion is not spoiled by ignoring it.
In figure~\ref{rms} we plot {\footnotesize$\sqrt{\left< b_{\perp}^2 \right>_n}$}
versus $\left<x\right>_n$ demonstrating that the transverse size of the nucleon again appears 
to depend significantly on the momentum fraction at which it is probed.  Note that 
$\left< b_{\perp}^2 \right>_2$ is not shown in the figure because $n=2$ corresponds
to the $q+\overline{q}$ distribution whereas $n=1,3$ correspond to the $q-\overline{q}$ distribution.
\subsection{$b_\perp$ Dependence}
\label{bdep}
Lattice calculations of parton distributions are restricted to a few low 
moments, however equation~\ref{mom} illustrates that the $\vec{b}_\perp$ dependence of the lowest
moments is quite illuminating.  For $n=1$ we have $\int\!dx\,q(x,\vec{b}_\perp)$
which is the transverse probability distribution without regard to the momentum fraction.
Furthermore the corresponding operator is conserved, hence $q(\vec{b}_\perp)$ is independent of the renormalization 
scale.  For $n=2$ we have $\int\!dx\,x\, q(x,\vec{b}_\perp)$ which is the transverse 
momentum distribution for which the sum over all quarks and gluons 
is again conserved and hence the total transverse momentum
distribution is also scale independent.  Unfortunately the moments $n\ge 3$ no longer have simple 
physical interpretations, nonetheless they 
represent the transverse distribution of the corresponding moment.

Assuming a dipole form for $A^q_{n0}$, which is justified by figure~\ref{AAA} for $Q^2 \le 3.5 \mathrm{GeV}^2$,
we can perform the integrals in equation~\ref{mom} and calculate the $\vec{b}_\perp$ dependence
of each fixed moment for ${\vec{b}_\perp}^{\,2} \ge (3.5 \mathrm{GeV}^2)^{-1}$. 
As figure~\ref{qqq} illustrates we again observe that the 
lowest moment is the most broadly distributed in the transverse plane and higher moments are 
successively more narrow.
\section{Conclusion}
\label{conc}
Our lattice calculations of the nucleon's generalized form factors have yielded 
insight into the quark structure of the nucleon.  We have demonstrated that 
for a world with heavy pion masses the quark orbital motion constitutes 
roughly $20\%$ of the nucleon's spin.  Additionally we have shown, in several ways,
that the transverse distribution of quarks within the nucleon depends
significantly on the momentum fraction of the quarks.  We are currently 
pursuing calculations with significantly lighter pion masses~\cite{Renner:2004ck}
in an effort to examine the quark substructure of the physical nucleon.
\section*{References}
\bibliography{drubrenner-ghp-04}
\bibliographystyle{h-physrev3}\nopagebreak
\end{document}